# Room-temperature Distributed Feedback CsPbBr$_3$ Perovskite Laser Integrated on a Silicon Nitride Waveguide Platform

Federico Fabrizi[1,2], Piotr J. Cegielski[1], Saeed Goudarzi[2], Naho Kurahashi[3], Manuel Runkel[3], Cedric Kreusel[3], Bartos Chmielak[1], Stephan Suckow[1], Thomas Riedl[3], Surendra B. Anantharaman[1,4], Maryam Mohammadi[1,*], Max C. Lemme[1,2,*]

[1]AMO GmbH, Otto-Blumenthal-Straße 25, Aachen, 52074, Germany
[2]RWTH Aachen University, Templergraben 55, Aachen, 52062, Germany
[3]Bergische Universität Wuppertal, Gaußstraße 20, Wuppertal, 42119, Germany
[4]Low-dimensional Semiconductors Lab, Department of Metallurgical and Materials Engineering, Indian Institute of Technology Madras, Chennai 600036, India
*Corresponding Authors: mohammadi@amo.de; max.lemme@eld.rwth-aachen.de

**Abstract**

Silicon photonic integrated circuits (PICs) require cost-effective laser sources that can be monolithically integrated. The low cost and low-temperature solution processability of metal halide perovskites (MHPs) make them attractive alternatives to established III-V compound semiconductors for on-chip laser sources in PICs. Cesium lead bromide (CsPbBr$_3$) perovskites are emerging materials for green light-emitting diodes and lasers. To date, amplified spontaneous emission (ASE) at room temperature has been frequently achieved in CsPbBr$_3$ thin films, while reports on lasing are more limited. Here, we demonstrate a first-order grating distributed feedback (DFB) CsPbBr$_3$ thin-film laser operating at room temperature. Planar hot-pressed (PHP)-CsPbBr$_3$, with a low ASE threshold of 14.5 µJcm$^{-2}$ under 0.3 nanosecond (ns) pump pulses, was monolithically integrated into a silicon nitride (Si$_3$N$_4$) waveguide platform via a compatible top-down patterning process. The first-order grating DFB PHP-CsPbBr$_3$ thin-film laser operated at 540 nm in the green spectral region, where III-V lasers have limitations, and exhibited a lasing threshold of 0.755 mJcm$^{-2}$ at room temperature. This work marks a significant step toward utilizing MHPs for on-chip green lasers in PICs for commercial applications.

**Introduction**

In the past decade, the commercialization of silicon photonics has accelerated, driven by substantial advancements in both photonic component performance and integration complexity [1]. Silicon enables low-loss waveguides [2], passive and active photonic components with high integration density [3], and high-yield mass fabrication [4], all because of its compatibility with complementary metal-oxide semiconductor (CMOS) manufacturing technology [5,6]. However, as an indirect semiconductor, silicon is not an efficient light source. Therefore, silicon is typically combined with III-V materials for light sources through a heterogeneous integration approach [7]. This approach involves the direct bonding of well-established III-V light-emitting sources onto silicon photonic integrated circuits (PICs). Bonding techniques incur substantial additional costs for packaging and assembly, along with significant increases in coupling losses [1]. Monolithic integration is another promising approach for integrating III-V lasers into PICs. This refers to the growth of III-V compound semiconductor materials directly on PICs via molecular beam epitaxy (MBE), chemical vapor deposition (CVD), or vapor phase epitaxy (VPE). The main challenges of monolithic integration include anti-phase domains (APDs), dislocations, and thermal cracks due to lattice mismatches between III-V and group IV materials [8,9].

Metal halide perovskites (MHPs) are a class of solution-processable ionic semiconductors with remarkable optoelectronic properties [10], such as carrier diffusion lengths above 1 µm [11–13], high photoluminescence (PL) quantum yields [14,15] and tunable bandgaps [16,17] that have shown promise for applications in solar cells and light-emitting diodes (LEDs) [18–22] and lasers [23–26]. Moreover, compared with established III-V compound semiconductors, MHPs do not require expensive epitaxial growth processes at high temperatures on crystalline substrates with matching lattice constants [27,28]. This offers significant advantages for monolithic integration directly onto silicon wafers [29,30]. They can also emit in the green spectrum at room temperature, where established III-V semiconductor gain media such as indium-gallium-nitride (InGaN) or aluminum-gallium-indium-phosphide (AlGaInP) are limited [31]. All-inorganic cesium lead bromide ($CsPbBr_3$) perovskite thin films have been investigated as a gain medium in PICs to address the green gap. However, early attempts to use solution-processed $CsPbBr_3$ thin films for green lasers resulted in amplified spontaneous emission (ASE) only at cryogenic temperatures below 150 K [32] and, more recently, at room temperature via a planar hot-pressing (PHP) process [33]. The PHP process has also proven effective in fabricating vertical-cavity surface-emitting lasers (VCSELs) at room temperature [33]. Here, we present a combination of the PHP process and precise top-down lithography[30,34] to integrate a thin film of $CsPbBr_3$ as a gain medium on a silicon nitride ($Si_3N_4$) waveguide. The significant difference in refractive indices between MHPs and typical waveguide materials, such as $Si_3N_4$, makes extracting light from the perovskite into waveguides challenging [35,36]. We address this challenge by incorporating an external first-order grating cavity within the $Si_3N_4$ waveguide. This external cavity provides the means to achieve room-temperature $CsPbBr_3$ thin-film first-order grating distributed feedback (DFB) on-chip lasing. Our developed integration technology with top-down lithography is compatible with very sensitive and chemically unstable MHPs and advances the manufacturability of $CsPbBr_3$ green lasers.

**Results and discussion**

We investigated the monolithic integration of $CsPbBr_3$ thin films as gain media in on-chip photonics by combining spin-coating, PHP, and a compatible photolithography process[30,34]. The $CsPbBr_3$ thin films were spin-coated from a solution of cesium bromide (CsBr) and lead bromide ($PbBr_2$) in dimethyl sulfoxide (DMSO) onto $SiO_2$ substrates (further details are provided in the Experimental section). The deposited layers were recrystallized through a PHP process at 150°C under a pressure of 100 bar, using a flat silicon wafer as a stamp (see the Experimental section). The PHP-$CsPbBr_3$ was patterned via our developed technology involving lithography and reactive ion etching (RIE) processes (see the Experimental section). To assess the impact of the process integration steps on the ASE of the $CsPbBr_3$, we evaluated the ASEs of the spin-coated $CsPbBr_3$, the PHP-$CsPbBr_3$, and the patterned PHP-$CsPbBr_3$ films. The spin-coated $CsPbBr_3$ thin films had grains smaller than 0.3 μm, contained pinholes, and exhibited a root mean square (RMS) roughness of 21.85 ± 2.51 nm (Figure 1a). These films did not show ASE at room temperature (see Support Information (SI) Figure S1), even under pulsed excitation at very high fluence levels (λ = 355 nm, pulse duration ≈ 0.3 ns). Upon recrystallization via the PHP process, the RMS roughness decreased to 0.87 ± 0.42 nm, and the average grain size increased to 1.20 ± 0.12 µm (Figure 1b). After the patterning process, the RMS roughness and average grain size of the PHP-CsPbBr$_3$ thin film did not significantly change and were 1.55 ± 0.37 nm and 1.31 ± 0.09 µm, respectively (Figure 1c). The PHP-CsPbBr$_3$ thin film demonstrated ASE at room temperature, with a spectral width of approximately 2.7 nm (Figure 1d) and a threshold of 14.5 µJcm$^{-2}$ (Figure 1e), which aligns with previously reported results [33]. The ASE threshold value further decreased to 8.9 µJcm$^{-2}$ (Figure 1e) after the patterning process, demonstrating that the nanofabrication process does not impair the ASE properties of the PHP-CsPbBr$_3$ films. Moreover, the phase and optical stability of the CsPbBr$_3$ thin films after the structuring process were examined via X-ray diffraction (XRD) measurements and PL. The XRD patterns of the pristine, pressed, and patterned films show characteristic peaks of the pure orthorhombic CsPbBr$_3$ phase (reference code: 00-018-0364) (Figure 1f). More intense peaks are observed for the pressed and patterned films, which results from an overall improved crystallinity. Additionally, PL emission maps over scan areas of 25 µm × 25 µm (Figures 1g, h, and i), where the color scale indicates the emission peak wavelength position, show that the emission wavelength is nearly the same, 522 nm, for all samples. On average, pristine CsPbBr$_3$ shows green emission at 522.35 ± 0.15 nm, PHP-CsPbBr$_3$ before patterning at 522.71 ± 0.75 nm and PHP-CsPbBr$_3$ after patterning at 521.90 ± 0.29 nm. The full width at half maximum (FWHM) of the PL spectrum did not change after the PHP and patterning processes in the same scanning area, with an average value of approximately 14.5 nm (SI Figures S2a-c). These findings prove that the crystallinity and optical properties of the CsPbBr$_3$ thin films remain stable throughout the patterning processes.

Next, the PHP-CsPbBr$_3$ was monolithically integrated into the Si$_3$N$_4$ photonics chips. We deposited the perovskite onto a quarter wavelength shifted (QWS) first-order grating that was etched into a Si$_3$N$_4$ waveguide and cladded with a 50 nm SiO$_2$ planarization layer (see the Experimental section). A schematic of the DFB-integrated perovskite laser and its side view are shown in Figures 2a and 2b, respectively. The Si$_3$N$_4$ waveguide can support both transverse electric (TE) and transverse magnetic (TM) fundamental modes. The first-order grating allows in-plane emission into the waveguide and maximizes the effective refractive index of the waveguide mode. To achieve an overlap between the gain spectrum of PHP-CsPbBr$_3$ and the resonance wavelength of the DFB, the grating period was determined according to the Bragg condition: $\lambda_B = 2\Lambda n_{eff}/m$, where $\lambda_B$ is the resonance wavelength, $\Lambda$ is the grating period, $n_{eff}$ is the effective refractive index and m is the grating order, which is equal to 1. The refractive index of the PHP-CsPbBr$_3$ thin film was extracted from ellipsometry measurements and found to be approximately 2.4 at 522 nm (SI Figure S3). The effective refractive index, $n_{eff}$, was calculated directly from the Lumerical software. $\Lambda$ was designed with variations ranging from 134 nm to 141 nm to accommodate for fluctuations in the perovskite refractive index. AFM scans of the fabricated first-order gratings etched into the waveguides are shown in SI Figures S4a and b. First-order grating simulations with etching depths of 25 nm and 50 nm were conducted (SI Figures S5a and b). The grating with an etch depth of 25 nm minimizes scattering losses in the distributed resonator. In contrast, asymmetry becomes evident owing to the high loss in the reflection coefficient for the grating with an etch depth of 50 nm. The optimal thickness of the PHP-CsPbBr$_3$ gain medium was estimated through finite difference time domain (FDTD) simulations via commercially available software (Lumerical FDTD Solutions) (SI Figure S6). Owing to the refractive index contrast, the optical mode is fully confined in the PHP-CsPbBr$_3$ layer when its thickness is ≥ 110 nm, at which point the perovskite acts as a waveguide (SI Figure S6). The mode overlap with the Si$_3$N$_4$ waveguide becomes significant for perovskite thicknesses ≤ 70 nm. However, achieving a uniform thin layer of CsPbBr$_3$ through a solution process is challenging. Here, the pristine CsPbBr$_3$ thickness was approximately 120 nm. After the PHP process, it was approximately 90 nm, which is the minimum thickness that we can technologically achieve. For the PHP-CsPbBr$_3$ composite with a thickness of 90 nm, the mode overlap between the Si$_3$N$_4$ waveguide and the perovskite film is approximately 24% (Figure 2c). The grating with an etch depth of 25 nm, in combination with the relatively low thickness of the perovskite film of 90 nm, ensured that the mode was guided mainly in Si$_3$N$_4$, allowing interaction with the perovskite to occur via the evanescent field of that mode.

A schematic of the PHP-CsPbBr$_3$ slab integration process on Si$_3$N$_4$ PICs via photolithography with a double-layer resist is shown in Figure 3a. The pattern was etched through a single-step process via inductively coupled plasma reactive ion etching (ICP RIE) using HBr and BCl$_3$ gases, which removed the resist interlayer and the exposed CsPbBr$_3$ film (see the Experimental section for details). The PHP-CsPbBr$_3$ layers, each 90 nm thick (SI Figure S7), were placed on top of a single-mode Si$_3$N$_4$ rib waveguide, cladded with SiO$_2$. Optical and scanning electron microscopy (SEM) images of the laser devices made from the etched PHP-CsPbBr$_3$ film after resist stripping are shown in Figures 3b and 3c, respectively. We characterized the patterned PHP-CsPbBr$_3$ on top of the waveguides with scanning PL spectroscopy. The PL peak position (Figure 3d) and the FWHM of the PHP-CsPbBr$_3$ slab (SI Figure S8) after integration into the Si$_3$N$_4$ waveguides are consistent with those of pristine CsPbBr$_3$, around 522 nm and 14.5 nm, respectively.

The monolithically integrated PHP-CsPbBr$_3$ laser was pumped from the top with ns laser pulses at a wavelength of 355 nm. The spectra of the PHP-CsPbBr$_3$ lasers, averaged over 10 s, were recorded using a multimode optical fiber connected to a spectrometer. Figure 4a illustrates the schematic of the characterization setup. Figure 4b shows the evolution of the emission spectra from the monolithically integrated PHP-CsPbBr$_3$ laser over a range of pump fluences at room temperature. A wide peak at approximately 522 nm was observed at low excitation fluence, which is related to the PL of the PHP-CsPbBr$_3$ thin film. As the pump power was increased to 0.775 mJcm$^{-2}$, single-mode lasing appeared at the longer-wavelength shoulder of the photoluminescence spectrum for the grating with a period of 138 nm. The FWHM decreased from approximately 14.5 nm to 0.5 nm, a clear indication of lasing (Figure 4c). In Figure 4d, the FWHM and output intensity are depicted as a function of the pump fluence. The output power exhibits a superlinear increase with increasing pump power, along with a pronounced narrowing of the mode linewidth at 0.839 mJcm$^{-2}$ and 0.935 mJcm$^{-2}$. The threshold level was determined by the intersection of the two linear fits below and above the threshold emissions, yielding a threshold level of 0.755 mJcm$^{-2}$. Importantly, in this work, threshold measurements were performed via ns pulses, which could increase the lasing threshold compared to femtosecond (fs) pulses[37].

**Conclusions**

A CsPbBr$_3$ DFB laser operating at room temperature and fully monolithically integrated into a silicon nitride waveguide platform was demonstrated. A PHP-CsPbBr$_3$ film with a low-threshold ASE of 14.5 µJcm$^{-2}$ at room temperature was integrated into PICs via a top-down patterning process. This process is compatible with the CsPbBr$_3$ and with conventional silicon process technology. The light produced by CsPbBr$_3$ was transferred to the Si$_3$N$_4$ waveguides through a first-order grating cavity and amplified through evanescent field interactions with the perovskite gain medium. The first-order grating enabled in-plane emission into the waveguide and maximized the effective refractive index of the waveguide mode, presenting a novel approach to optimizing light coupling. The CsPbBr$_3$ thin-film DFB laser exhibited a threshold of 0.755 mJcm$^{-2}$ (0.3 ns pump pulses) at room temperature. This work highlights the potential of CsPbBr$_3$ for PIC-integrated green lasers and addresses the green color gap in III–V semiconductor laser technology.

**Experimental Section**

**Perovskite Deposition**

The CsPbBr$_3$ perovskite solution was prepared following a previously published recipe[33]. The two precursors, cesium bromide (CsBr, 99.999%, trace metal basis, Sigma Aldrich) and lead bromide (PbBr$_2$, 99.999%, ultradry, Alfa Aesar), were mixed in anhydrous dimethylsulfoxide (DMSO, anhydrous, ≥99.9%). The molar ratio of CsBr:PbBr$_2$ was 2.75:1 for all the samples. The solution was stirred and heated overnight at 60°C in a nitrogen-filled glovebox and filtered through a 0.2 µm filter before use. The perovskite layer was prepared by spin-coating the precursor solution at 4000 rpm, 120 s, and 11 s. After spin-coating, the film was annealed on a hotplate at 100°C for 20 min. A thermal imprint process, termed 'planar hot-pressed', was employed to recrystallize and increase the grain size of the as-deposited layer at 150°C and 100 bar pressure, with a flat silicon wafer used as the stamp.

**Patterning of CsPbBr$_3$**

A compatible top-down patterning process was used to integrate the CsPbBr$_3$ layer into the Si$_3$N$_4$ photonic platform, as described in our previous work[30,34]. First, a 200 nm poly (methyl methacrylate) (PMMA) resist layer was deposited via spin-coating. AZ MiR 701 positive tone photoresist was then spin-coated at 3000 rpm and annealed at 95°C for 90 s. Next, the area of the active laser on top of the first grating was defined through a mask by ultraviolet (UV) light using a contact lithography mask aligner (EVG 420). Afterwards, the photoresist was baked at 115°C for 90 s and developed for 33 s in an MF26A CD developer. The pattern was transferred into the CsPbBr$_3$ with a single reactive ion etching step with HBr and BCl$_3$ gases. The photoresist was removed by dipping the samples in chlorobenzene at 60°C for 30 min. Finally, 1 µm of PMMA was spin-coated and annealed at 80°C for 10 min as a protective layer.

**Fabrication of the PICs**

The PICs were fabricated on 150 mm silicon wafers with 2.2 µm of thermally grown SiO$_2$. A total of 400 nm of silicon nitride was deposited via a low-pressure chemical vapor deposition (LPCVD) process from dichlorosilane and ammonia gases. The gratings were patterned via electron beam lithography (Vistec EBPG 5200) and etched via inductively coupled plasma reactive ion etching (ICP-RIE, Oxford Instruments, PlasmaLab System 100) with CHF$_3$ and helium gases. This step was performed first on perfectly flat wafers to ensure a high-quality lithography process. As no etch stop layer was present, etching was performed in steps to control the etch depth with AFM. Next, the silicon nitride waveguides were patterned by optical lithography with a Canon i-line stepper and etched in an ICP-RIE reactor using CHF$_3$ and helium gases (Oxford Instruments, PlasmaLab System 100). This was followed by LPCVD deposition of 600 nm of SiO$_2$ (low-temperature oxide-LTO) via a silane and oxygen reaction. Planarization of the SiO$_2$ surface was performed by spin-coating spin-on glass (SOG) and subsequent etching in CHF$_3$ plasma. This process cycle of SOG coating and dry etching was repeated twice, resulting in an approximately 50 nm height difference between the LTO surface above and next to the silicon nitride waveguides with a smooth transition. Approximately 300 nm of LTO were consumed during these planarization steps, leaving 50 nm of LTO above the silicon nitride waveguides. Next, the wafers were coated with a protective polymer and diced into 3 cm × 3 cm samples.

**Photoluminescence Mapping**

Photoluminescence mapping was performed via a WiTec confocal Raman microscope (model alpha300R). The sample was scanned with a 457 nm CW laser at 100 nW power. The laser spot was approximately 300 nm in diameter, and the scanning step size was 400 nm. The measurements were taken at room temperature under ambient conditions.

**AFM, XRD, and SEM inspection**

AFM images were taken via an atomic force microscope (Bruker Dimension Icon) in tapping mode. XRD with filtered Cu-Kα radiation (wavelength at 1.5405 Å) was carried out via a PANalytical instrument at a current of 40 mA and a voltage of 40 kV. SEM micrographs were taken with an SEM Zeiss SUPRA 60 at 4 kV and a working distance of 3.5 mm.

**Ellipsometry**

Spectroscopic ellipsometry of the perovskite thin films was conducted via a J. A. Woollam M-2000 spectroscopic ellipsometer. The acquired data were analyzed and modeled via the J. A. Woollam CompleteEASE software package.

**Laser Characterization**

The integrated perovskite laser was pumped from the top by a frequency-tripled diode-pumped solid-state laser (PowerChip NanoLaser, TEEM Photonics, France) with 0.3 ns laser pulses at a 355 nm wavelength and a 1 kHz repetition rate. The excitation spot was 350 µm. The excitation density was varied by a neutral density filter wheel. The power of the pump laser was measured with a thermal sensor head (S470C, Thorlabs). The sample was positioned on a sample chuck mounted on an XY micrometer stage and held under vacuum. To enable alignment of the optical fiber end to the sample edge, a microscope equipped with a 20x objective was positioned above the sample. The output was collected by end-fire coupling to a step-index multimode fiber. The multimode fiber was used to relax alignment tolerances because the samples were cleaved by hand after perovskite deposition, resulting in uneven edges. The spectra were measured with a spectrometer (Princeton Instruments, Acton SP2500; gratings: 300 and 1200 lines mm$^{-1}$).

**Acknowledgements**

This project has received funding from the European Union's Horizon 2020 research and innovation programme under the Marie Skłodowska-Curie grant agreement No 956270, the project FOXES (951774), the German Research Foundation (DFG) through the project Hiper-Lase (GI 1145/4-1, LE 2440/12-1 and RI1551/18-1) and the German Ministry of Education and Research (BMBF) through the project NEPOMUQ (13N17112 and 13N17113). The authors thank P. Grewe and Dr. U. Böttger (Electronic Material Research Lab, RWTH Aachen University) for their support in the XRD measurements.

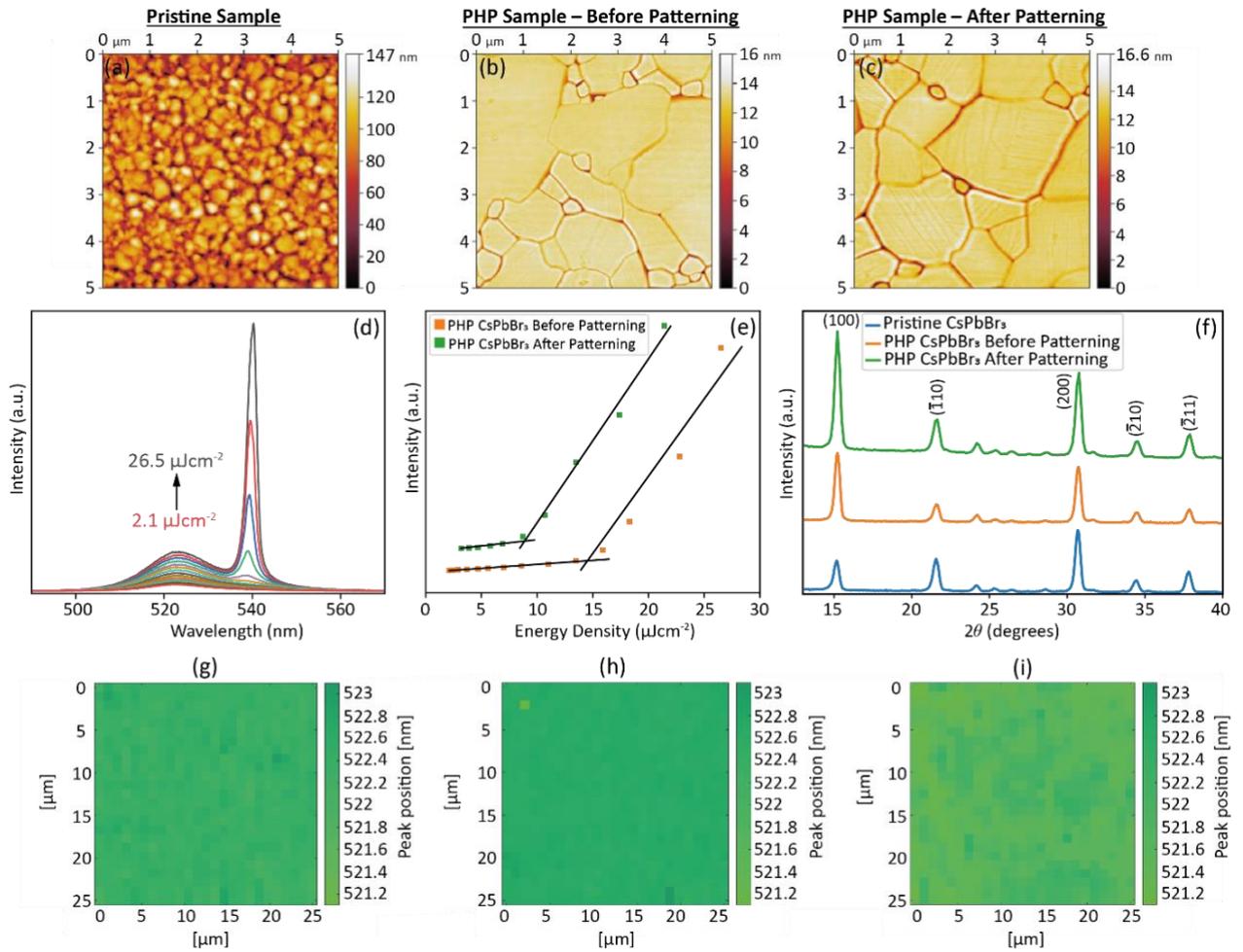

**Figure 1**. AFM scans of (a) the pristine sample, (b) after the PHP and (c) after the PHP and patterning processes on a 5 μm × 5 μm area. A clear increase in the grain size from the nm to the μm scale is visible after the planar hot process (PHP). (d) ASE spectrum of the PHP CsPbBr₃ thin film. (e) ASE threshold values for PHP CsPbBr₃ before and after the patterning process. (f) XRD spectra of the pristine sample, after the PHP process and after the PHP and patterning processes. Large-area scan of the peak position of (g) the pristine sample, (h) after the PHP and (i) after the PHP and patterning process.

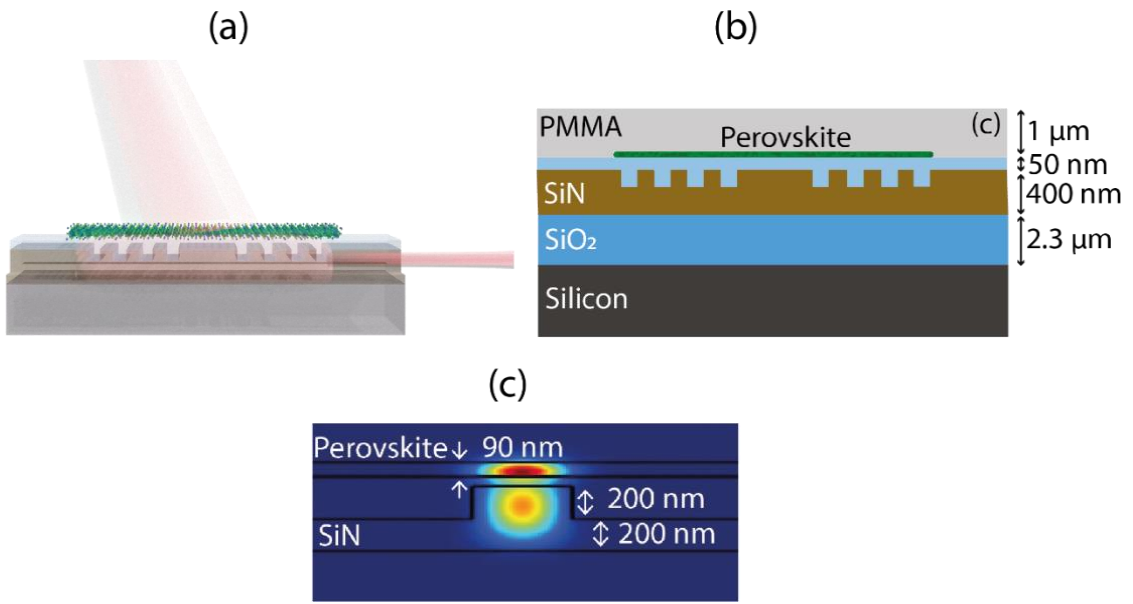

**Figure 2**. (a) 3D sketch of a CsPbBr$_3$ laser integrated on a standard silicon nitride photonic chip. (b) Side view of the DFB perovskite laser. (c) Simulated TE waveguide mode with a perovskite film thickness of 90 nm and a 50 nm SiO$_2$ spacer. The mode is guided mainly in the Si$_3$N$_4$ waveguide.

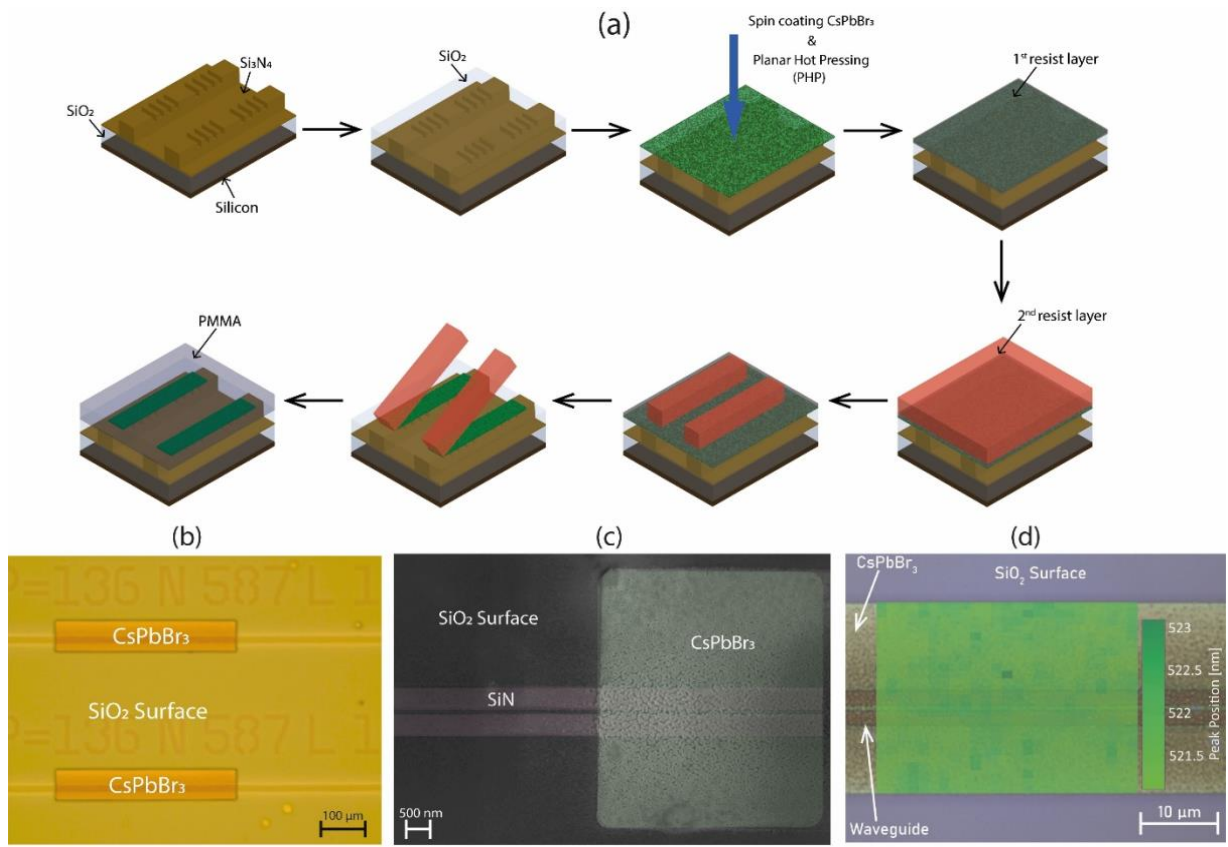

**Figure 3**. (a) Fabrication process flow (consequent steps marked with progressive arrows): Si$_3$N$_4$ deposition and patterning → deposition of CsPbBr$_3$ and planar hot pressing (PHP) → deposition of the bottom resist layer → deposition of the top photoactive resist layer → development of the second resist layer → pattern transfer by reactive ion etching and resist stripping by dissolving the bottom resist layer and lifting the top layer → deposition of 1 μm thick PMMA cladding. (b) Optical image of a final laser device after resist stripping. (c) SEM image of the patterned PHP-CsPbBr$_3$ film after resist stripping. The perovskite layer remains only on top of the grating, which forms the active laser area. (d) Overlay of an optical image of the final perovskite laser and a PL map of 30 μm X 35 μm. The peak position emission of CsPbBr$_3$ after etching on top of the Si$_3$N$_4$ waveguide is very uniform.

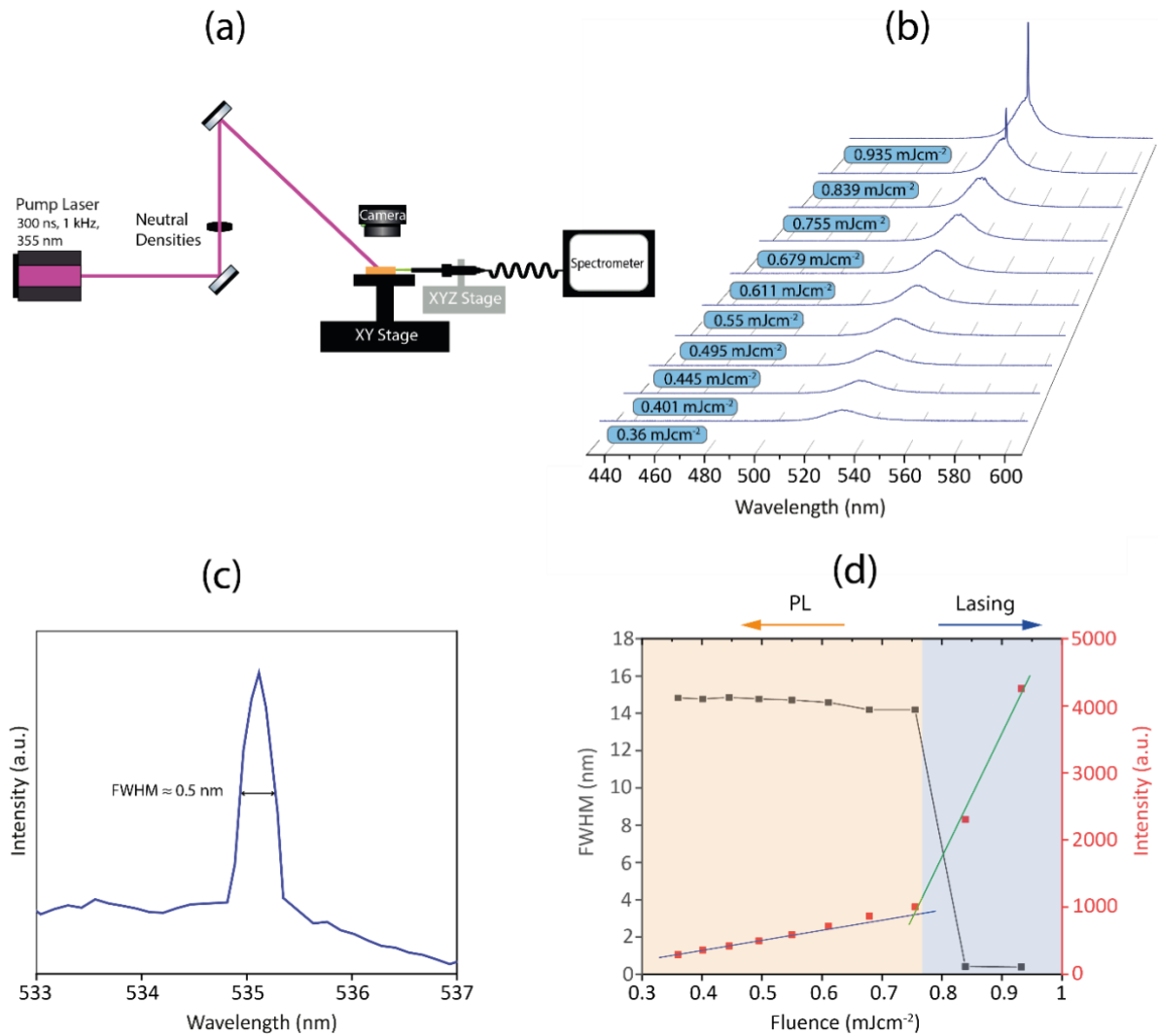

**Figure 4**. (a) Schematic illustration of the measurement setup. (b) Evolution of the emission spectrum with increasing excitation intensity from a broadband photoluminescence (PL) spectrum (FWHM ~ 15 nm) to a narrowband laser peak (FWHM ~ 0.5 nm). (c) A high-resolution spectrum of the DFB laser emission line. (d) Dependence of the FWHM (grey) and output intensity (red) on the pump fluence.

**SUPPORTING INFORMATION**

# Room-temperature Distributed Feedback CsPbBr$_3$ Perovskite Laser Integrated on a Silicon Nitride Waveguide Platform

Federico Fabrizi[1,2], Piotr J. Cegielski[1], Saeed Goudarzi[2], Naho Kurahashi[3], Manuel Runkel[3], Cedric Kreusel[3], Bartos Chmielak[1], Stephan Suckow[1], Thomas Riedl[3], Surendra B. Anantharaman[1,4], Maryam Mohammadi[1,*], Max C. Lemme[1,2,*]

[1]AMO GmbH, Otto-Blumenthal-Straße 25, Aachen, 52074, Germany
[2]RWTH Aachen University, Templergraben 55, Aachen, 52062, Germany
[3]Bergische Universität Wuppertal, Gaußstraße 20, Wuppertal, 42119, Germany
[4]Low-dimensional Semiconductors Lab, Department of Metallurgical and Materials Engineering, Indian Institute of Technology Madras, Chennai 600036, India
*Corresponding Authors: mohammadi@amo.de; max.lemme@eld.rwth-aachen.de

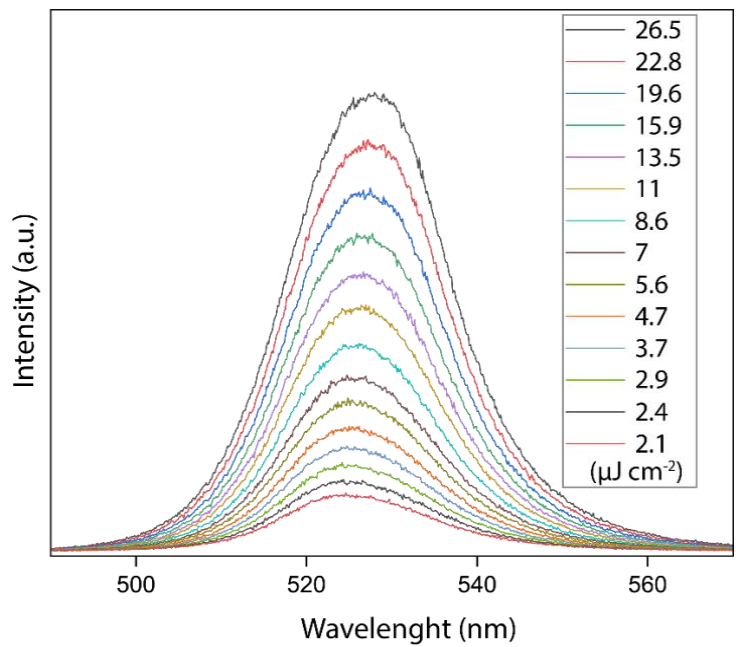

**Figure S1**. Photoluminescence of unpressed $CsPbBr_3$ at different excitation powers.

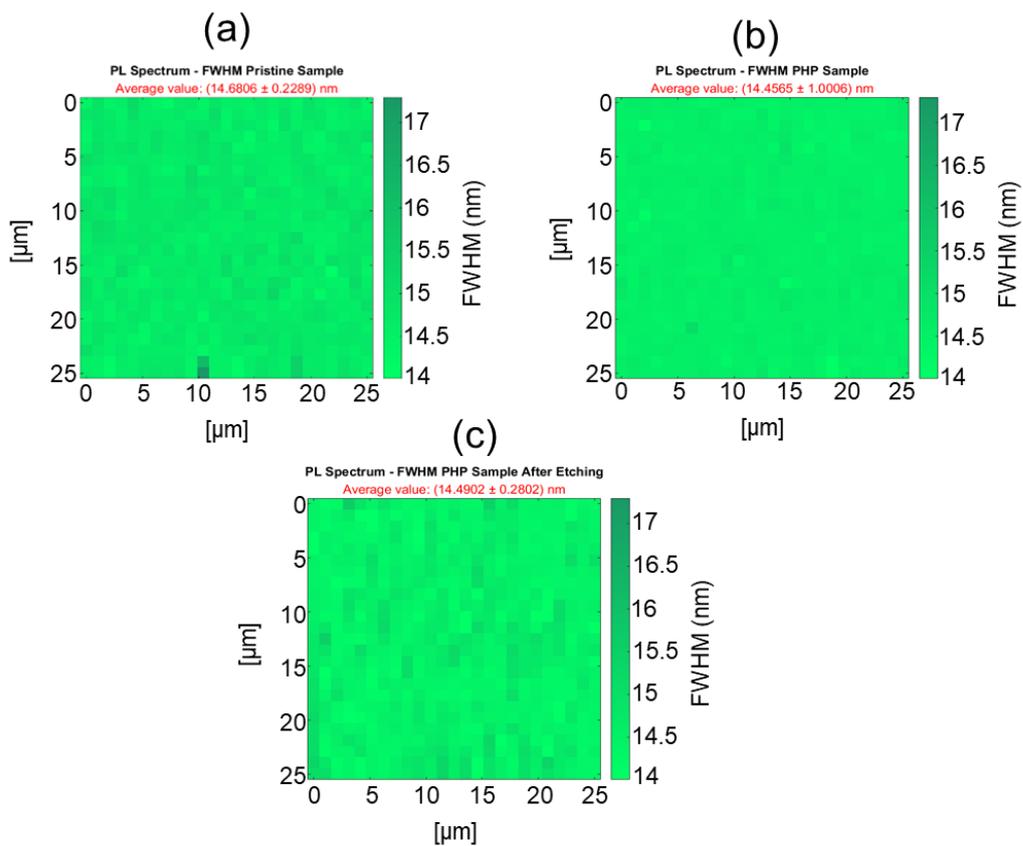

**Figure S2**. Large are scans of 25 μm x 25 μm of (a) the pristine $CsPbBr_3$, (b) after the PHP and (c) after the patterning process. Each pixel corresponds to a fitted value of the FWHM of the PL spectrum.

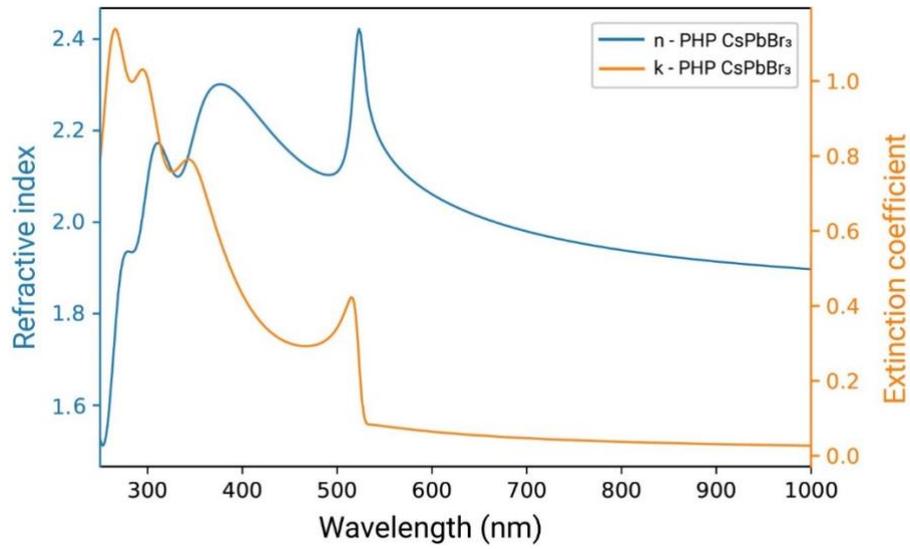

**Figure S3**. Refractive index (n) and excition coefficient (k) of PHP-CsPbBr$_3$ thin film.

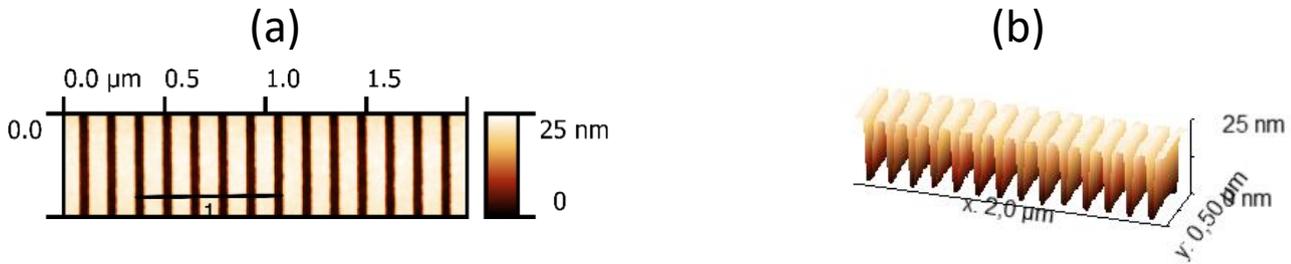

**Figure S4. (a)-(b)** AFM scan of the fabricated first-order gratings etched into the waveguides.

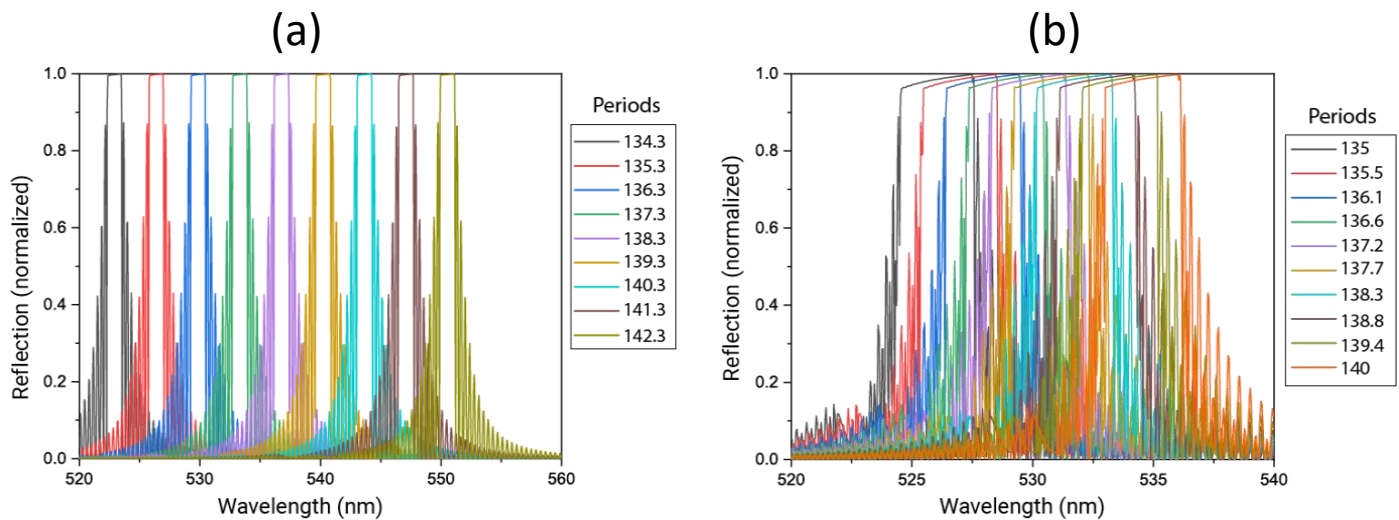

**Figure S5**. First-order grating simulations with an etching depth of (a) 25 nm and (b) 50 nm.

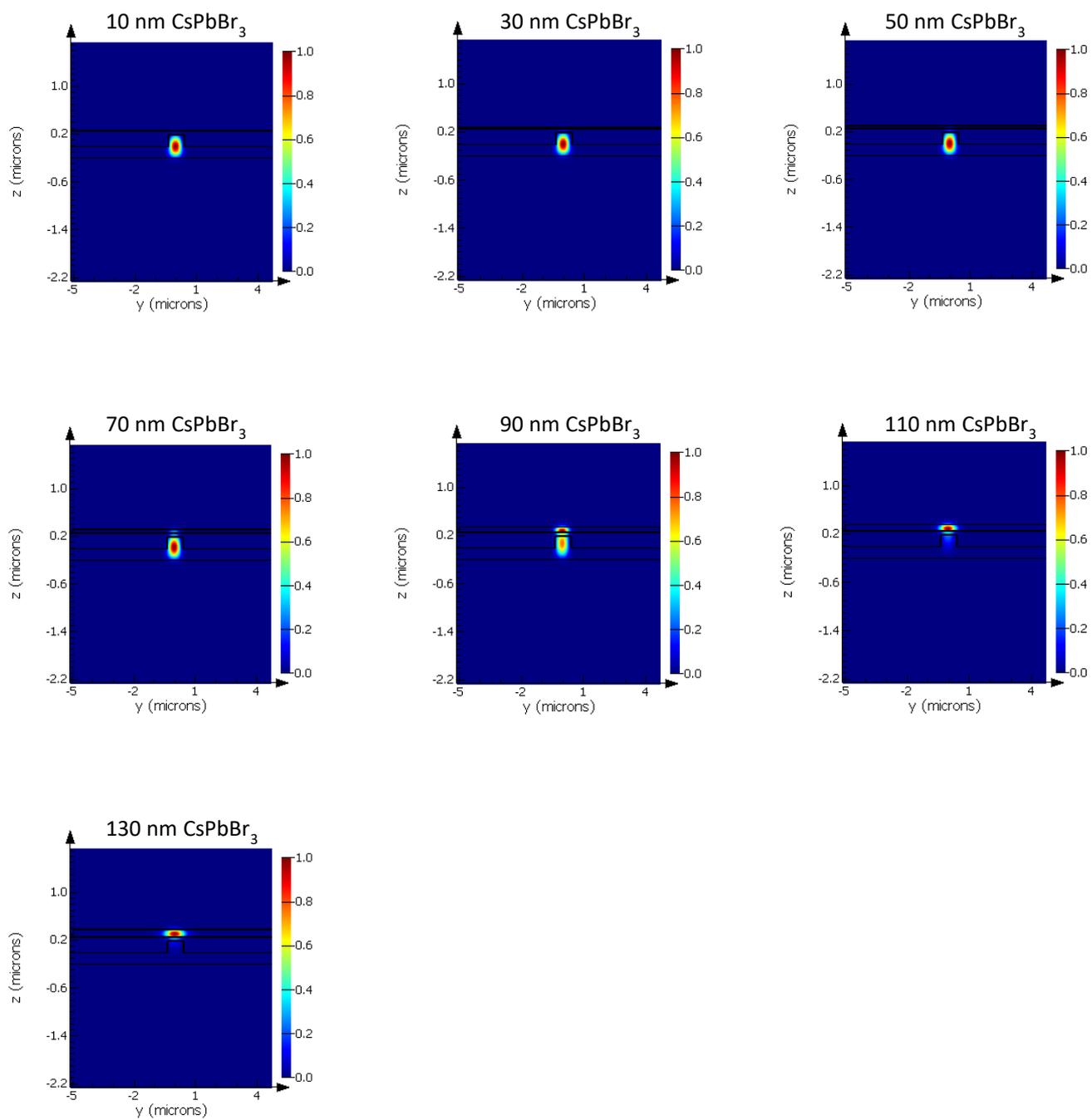

**Figure S6.** Optical mode evolution with different PHP-CsPbBr$_3$ thicknesses, from 10 nm to 130 nm.

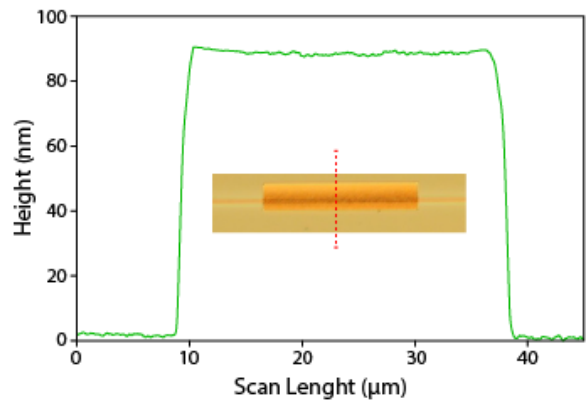

**Figure S7**. Profilometer of the PHP-CsPbBr₃ thin film on top of the waveguide after the patterning process.

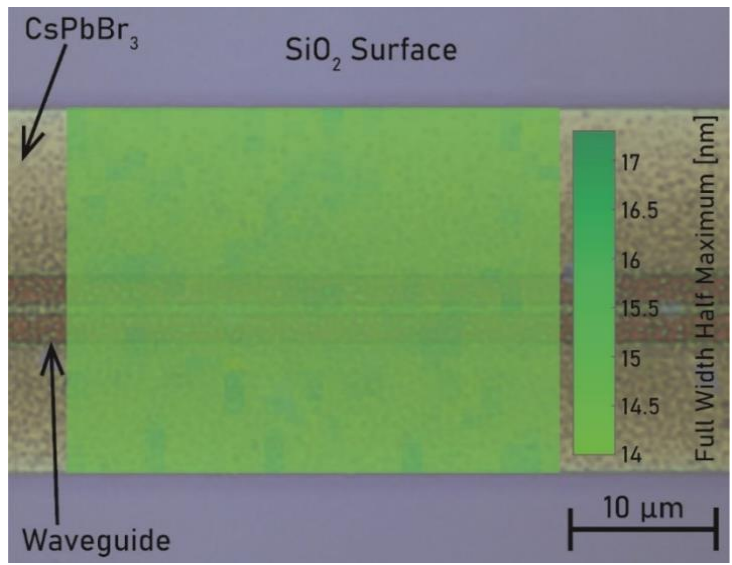

**Figure S8.** Large area PL scan of 30 μm X 35 μm FWHM of the patterned PHP-CspbBr₃ on top of the Si₃N₄ waveguide**.**